\documentclass[12pt,preprint]{emulateapj}
\usepackage{natbib}
\usepackage{graphicx,natbib}

\shortauthors{Miniati, Koushiappas \& Di Matteo}

\newcommand{\gr}{$\gamma$-ray }
\newcommand{\app}{\rm Astropart.~Phys.\ }
\newcommand{\cpc}{\rm Comp.\ Phys.\ Comm.\ }

\begin{document}
\title{Angular Anisotropies in the Cosmic Gamma-ray Background as a Probe of its Origin}

\author{Francesco Miniati\altaffilmark{1}, Savvas M. Koushiappas\altaffilmark{2}, 
Tiziana Di Matteo\altaffilmark{3}}

\altaffiltext{1}{Physics Department, Wolfgang-Pauli-Strasse 16, 
ETH-Z\"urich, CH-8093, Z\"urich, Switzerland; fm@phys.ethz.ch}

\altaffiltext{2}{Theoretical \& ISR Divisions, Los Alamos National Laboratory, P.O. Box 1663, 
Los Alamos, NM 87545, USA; smkoush@lanl.gov}

\altaffiltext{3}{Department of Physics, Carnegie-Mellon University, 5000 Forbes Ave., 
Pittsburgh, PA 15213, USA; tiziana@phys.cmu.edu}

\begin{abstract}
  Notwithstanding the advent of the {\it Gamma-ray Large Area
    Telescope}, theoretical models predict that a significant fraction
  of the cosmic \gr background (CGB), at the level of 20\% of the
  currently measured value, will remain unresolved.  The angular power
  spectrum of intensity fluctuations of the CGB contains information
  on its origin.  We show that probing the latter from a few tens of
  arcmin to several degree scales, together with complementary GLAST
  observations of \gr emission from galaxy clusters and the blazars
  luminosity function, can discriminate between a background that
  originates from unresolved blazars or cosmic rays accelerated at
  structure formation shocks.
\end{abstract}


\section{INTRODUCTION}

The existence of diffuse $\gamma$-ray radiation was established in the
early 70's at the dawn of $\gamma$-ray astronomy by the SAS 2 mission
~\citep{fichteletal75} and further assessed by COS B~\citep{bennet90}.
The determination of an extragalactic component (heretofore, CGB for
cosmic \gr background), which relies on a model for the foreground
Galactic emission, became possible only later with the advent of
the Compton Gamma Ray Observatory.  At energies above 30 MeV the EGRET
experiment onboard the CGRO measured a CGB characterized by a
specific intensity $I_\nu(\varepsilon)$=$ 5.4 (\varepsilon/{\rm
  keV})^{-2.1}$ ph cm$^{-2}$ s$^{-1}$ keV$^{-1}$ sr$^{-1}$ or an
integral intensity above 100 MeV, $I$=$1.45\times 10^{-5}$ ph cm$^{-2}$
s$^{-1}$ sr$^{-1}$~\citep{sreeku98}.  A recent reassessment of this
measurement changes the spectrum from a straight power law to a
convex one and reduces the integral intensity to $1.15\times
10^{-5}$ ph cm$^{-2}$ s$^{-1}$ sr$^{-1}$~\citep[][but see also Keshet
et al.~\citeyear{keshetetal04}]{stmore04}.

The origin of the CGB is still subject of debate.  One naturally
expects a population of unresolved blazars, a class of AGNs known to
emit up to the highest energies, to contribute to
it~\citep{pgfc93,stsa96,muepo00}.  Quantitative estimates, however,
remain uncertain and various approaches suggest a contribution of
order 25-50\% of the CGB~\citep{chmu98,nato06}.  Another candidate
mechanism is inverse Compton (IC) emission on CMB photons from
cosmic-ray (CR) electrons accelerated at structure formation shocks
both around galaxy clusters (GC) and cosmic filaments~\citep{lowa00}.
This process can contribute up to a fraction 20\% of the CGB, without
violating the existing EGRET upper limits on the \gr emission from
individual GCs~\citep{min02}.  Normal and/or starburst galaxies have
also been proposed as substantial \gr emitters~\citep{pabr02,tqw07}.
Contributions from $\pi^0$-decay due to CR protons in the core of
GCs~\citep[e.g.,][]{min02,cobl98} and annihilation of dark
matter are also possible~\citep[e.g.,][]{ando06}.  Note, however, that
except for blazars none of these proposed sources has been detected in
$\gamma$-rays as yet.

The Large Area Telescope onboard
GLAST\footnote{http://glast.gsfc.nasa.gov/} will advance our
understanding of the origin of the CGB, by improving our estimate of
the Galactic emission, by resolving part of the extragalactic
component due known \gr populations and, possibly, by detecting new
types of \gr sources.  According to current theoretical predictions, a
significant fraction ($\geq 25\%$) of the CGB will remain unresolved
by GLAST.  In this {\it Letter} we study the angular intensity
fluctuations of this residual component as a means to
discriminate among viable models for its origin. We consider the case
of blazars and CRs accelerated at structure formation shocks.
We use numerical simulations to model the relevant quantities involved
in the calculations, particularly the correlation function of the
sources. We predict a substantial difference in the
level of fluctuations in the two scenarios above, testable with GLAST from a
few tens of arcmin to several degrees.  This experiment, together with
improved determinations of the blazar luminosity function and \gr
emission from GCs by GLAST, will provide a discriminatory test for the
unresolved CGB in the GLAST era.  CGB fluctuations due to structure
shocks were also computed in~\cite{walo00} analytically, and for an
emission sufficient to account for the whole EGRET CGB.  Our work is
complementary to that of Ando et al. who, with different methods,
studied the fluctuations produced by (a) sources similar to those
considered here but resolved by GLAST~\citeyearpar{aknt06a}, and (b)
unresolved emission from blazars and annihilation in dark matter
halos~\citeyearpar{aknt06b}.  Finally, CGB fluctuations were also
studied by~\cite{zhabe04} for MeV emission from SN Ia, and
by~\cite{cuoco06} for nearby emission above $\geq$100 GeV.  Note that
the success of any of these experiments hinges on the ability to
assess and, if necessary, separate out fluctuations of Galactic
origin.  This requires a study the luminosity function and
statistical spatial distribution of Galactic $\gamma$-ray sources
similar to the one carried out below for extragalactic sources.

Throughout we assume a $\Lambda$CDM cosmological model with
$\Omega_m$=0.3, $\Omega_{\Lambda}$=0.7, $\Omega_b$=0.04, and $h\equiv
H_0/100\,{\rm km}\,{\rm s}^{-1} {\rm Mpc}^{-1}$=$0.7$
\citep{spergeletal03}.

\section{Model}

Given the integral intensity $I(\hat v)$, above
photon energy $\varepsilon_{0}$ and along a direction $\hat v$, the angular
two point correlation function characterizing the fluctuation $\delta
I(\hat v)\equiv I(\hat v) - \langle I \rangle_\Omega $ about the
solid-angle-average value $\langle I \rangle_\Omega$, is the ensemble
average $ C(\theta) $=$ \langle \delta I(\hat v) \delta I(\hat u)
\rangle$, where $\theta$=$\arccos(\hat u\cdot\hat v)$.  When the
correlation length of the sources is small compared to the
spatial-scale over which they evolve and for small
angular separations, $\theta \ll $1 radian, we can
write~\citep{limber53,peebles93}
\begin{equation} 
C(\theta) =
\frac{1}{(4\pi)^2} \!
\int_{z_{min}}^{z_{max}} \! dz
 \frac{dV}{dz}\frac{\langle j(z)\rangle_V^2}{4\pi d_L^2}
\! \int_{-\infty}^{\infty} \! dy \, \xi(r[y,z]),
\label{limb:eq}
\end{equation} 
where
$\langle j(z) \rangle_V$ is the redshift dependent photon emissivity
of unresolved sources in units ph$\,$s$^{-1}\,$cm$^{-3}$ and averaged
over the comoving volume $V$, and $C(\theta)$ is in units
ph$^2\,$s$^{-2}\,$cm$^{-4}\,$sr$^{-2}$.  The source correlation
function is $\xi(r)$, $d_L$ is the luminosity distance and
$r(y,z)$=$(y^2+d_C^2 \theta^2)^{1/2}$, with $d_C$ the comoving
distance.  For sources characterized by a rest frame spectral
luminosity, $L_\nu(\varepsilon)$, and a comoving luminosity function,
${\cal L}_\gamma(L_\gamma,z)$, with $L_\gamma$=$\nu L_\nu$ computed at
$\varepsilon_0$ in units erg s$^{-1}$, we have
\begin{equation} 
\langle j(z) \rangle = 
\int_{0}^{L_{lim}(z)} 
dL_\gamma\,{\cal L}_\gamma(L_\gamma,z)\, L_{ph}(>\!\varepsilon_0[1+z]),
\label{jave:eq}
\end{equation}
where, $L_{ph}(>\!\varepsilon_0)$=$\int_{\varepsilon_{0}}^\infty
d\varepsilon\,\varepsilon^{-1} L_\nu(\varepsilon)$, is the integrated
photon luminosity above a threshold $\varepsilon_0$.  In
Eq.~(\ref{jave:eq}), $L_{lim}(z)$ is the luminosity of a source at
redshift $z$ with photon luminosity
$L_{ph}(>\!\varepsilon_0[1+z])$=$4\pi d_L^2(z)F_{lim}$, where
$F_{lim}(>\!\varepsilon_0)$ is the limiting integrated photon flux
above $\varepsilon_0$ where a source is resolved.  If the spectral
emissivity is a power law with index $\alpha$,
$L_\nu(\varepsilon)\!\propto\! \varepsilon^{-\alpha}$, then
$L_{lim}(z)=4\pi d_L^2 (\alpha
-1)(1+z)^{\alpha-2}\varepsilon_0F_{lim}$.
The angular power spectrum of the fluctuations is given by the
transform, $C_\ell$=$2\pi\int_0^\pi d({\rm cos}\theta) P_\ell({\rm
  cos}\theta) C(\theta)$, where $P_\ell$ is Legendre's polynomial of
order $\ell$.  Point sources also contribute Poisson noise to the
power spectrum and for unresolved sources this is~\citep{teef96}
\begin{equation} \label{cl1:eq}
C_\ell^{P}\! =\!
\frac{1}{(4\pi)^2}\!
\int_0^{z_{max}}\! dz 
\frac{dV/dz}{(4\pi d_L^2)^2} \!
\int_{0}^{L_{lim}(z)}\! dL_\gamma \,{\cal L}_\gamma 
L_{ph}^{2}.
\end{equation}
%

%
The average emissivity and correlation function of the CRs component
are obtained through a numerical simulation of structure formation
that follows the acceleration, spatial transport and energy losses of
CR particles.  This calculation uses the same technique as
in~\cite{min02}, but a volume eight times as large.  Shocks are
identified~\citep{minetal00} and a fraction of the particles crossing
them is accelerated to a power-law distribution in momentum space with
log-slope determined by the shock Mach number in the test particle
limit~\citep{bell78a}. The acceleration efficiency, based on a variant
of the thermal leakage prescription~\citep{kajo95,mrkj01}, when
expressed in terms of CR to shock ram pressure is always $< $40\% and
$< $1\% for protons and electrons, respectively.  The \gr emitting
particles are affected by energy losses, dominated by synchrotron and
IC emission for the CR electrons, and adiabatic losses, Coulomb and
inelastic p-p collisions for CR protons.  The latter process generates
{\it secondary} $e^\pm$ which are also followed by the simulation as
the primary electrons.  The simulation is carried out with the
cosmological code presented in~\cite{rokc93} in combination with the
CR code described in~\cite{min01, min02}.  We use a computational box
of 100 $h^{-1}$ Mpc on a side, with a grid of 1024$^3$ cells and
512$^3$ dark matter particles.  Momentum space is divided into five
log-bins.  We compute the emissivity as a function of cell position
and redshift, $j({\bf x},z)$, due to IC emission from CR electrons,
and due to the combined emission from $\pi^0$-decay and IC from
e$^\pm$.  We simply refer to the latter as hadronic. The volume
average is $\langle j(z)\rangle$=$V^{-1}\sum_{{\bf x}\in V} j({\bf
  x},z)$.  The two point correlation function is then given by,
$\xi(r,z)$=$(2\pi)^{-3}\int P_{\delta}(k,z) (\sin kr/kr) 4\pi k^2dk$,
with $P_{\delta}(k,z)$ the power spectrum of $\delta({\bf
  x},z)$=$j({\bf x},z)/\langle j(z)\rangle-1$.

%

%
We compute the correlation function of blazars in the following way.
We first use a high resolution N-body simulation and adopt a set of
physically and observationally motivated prescriptions to populate
collapsed halos with radio galaxies~\citep[see][for
details]{dicimi04}.  As illustrated in~\cite{dicimi04} the model of
radio galaxies accurately describes the observed angular correlation
function as well as the radio luminosity function up to redshift
$z\simeq$3. We rely on this model to compute the source correlation
function. We use the estimator, $
\xi(r,z)$=$(N^{ij}[r,z]/N^{ij}_{random}[r,z])-1$, where
$N_L(r,z)$
is the sum over pairs whose distance, $d_{ij}$, falls within $r$ and
$r$+$dr$, and $N^{ij}_{random}(r,z)$ is the same quantity but expected
for randomly distributed galaxies. We use this estimator instead of
one in which the pairs are weighted with the product of the pair
luminosity because for the angular scales relevant here it gives fully
consistent but much less noisy results~\citep{dicimi04}.

Following previous work, we describe the \gr luminosity function of
blazars, ${\cal L}_\gamma$, with the luminosity function of AGNs at
different wavelengths.  We consider two models: the Pure Luminosity
Evolution (PLE), based on the radio luminosity function, originally
proposed in~\cite{stsa96} and recently revised in~\citet{nato06} to
improve the redshift distribution of EGRET blazars; and the
Luminosity-Dependent Density Evolution (LDDE), based on the X-ray
luminosity function of AGN~\citep{has05}, and studied in detail
in~\citet{nato06}.  We set ${\cal L}_\gamma (L_\gamma,z)$=$ \kappa_s
(L_s/L_\gamma){\cal L}_s(L_s,z)$ where $s$= R, X indicates radio and
X-ray respectively, and assume a simple linear relation, $L_\gamma $=$
10^{q_s}L_{s}$, between the \gr and the radio/X-ray luminosity of the
sources.  The relevant parameters are $(\kappa_{\rm R},q_{\rm R})$=$
(0.173,3.28)$ and $(\kappa_{\rm X},q_{\rm X})$=$(5.11 \times 10^{-6},
3.80)$. The assumed $\gamma$-ray spectrum is a power law,
$L_\nu\propto\nu^{-\alpha}$, with $\alpha \approx 2.2$.  We use the
luminosity function thus built to estimate the unresolved average
emissivity from with Eq.~(\ref{jave:eq}).  Note that given the assumed
linear relations between radio, X-ray and \gr luminosity of the
sources, $\xi$ does not depend on wavelength, as expected in unified
AGN models in which the difference between blazars and radio galaxies
is only ascribed to the jet orientation with respect to the line of sight.

\section{Results and Discussion}
In the following we consider the integral photon intensity above 
$\varepsilon_0$=100 MeV and adopt a flux sensitivity for GLAST, 
$F_{lim}(>\!\varepsilon_0)$=$2\times 10^{-9}$ph s$^{-1}$cm$^{-2}$.
We provide a crude but robust lower limit for the unresolved CR
contribution by setting $z_{min}\!\approx \!(H_0/c)
(L^{max}_{ph}[>\!\varepsilon_0]/4\pi F_{lim})^{1/2}$ in
Eq.~(\ref{limb:eq}), where $L^{max}_{ph}(>\!\varepsilon_0)$ is the
luminosity of the brightest simulated GCs. This $z_{min}$ is the
redshift beyond which the brightest and, therefore, all simulated GCs
would be unresolved by GLAST even if they were point sources. We find
$z_{lim} \approx 0.05$.

\begin{figure}
\plotone{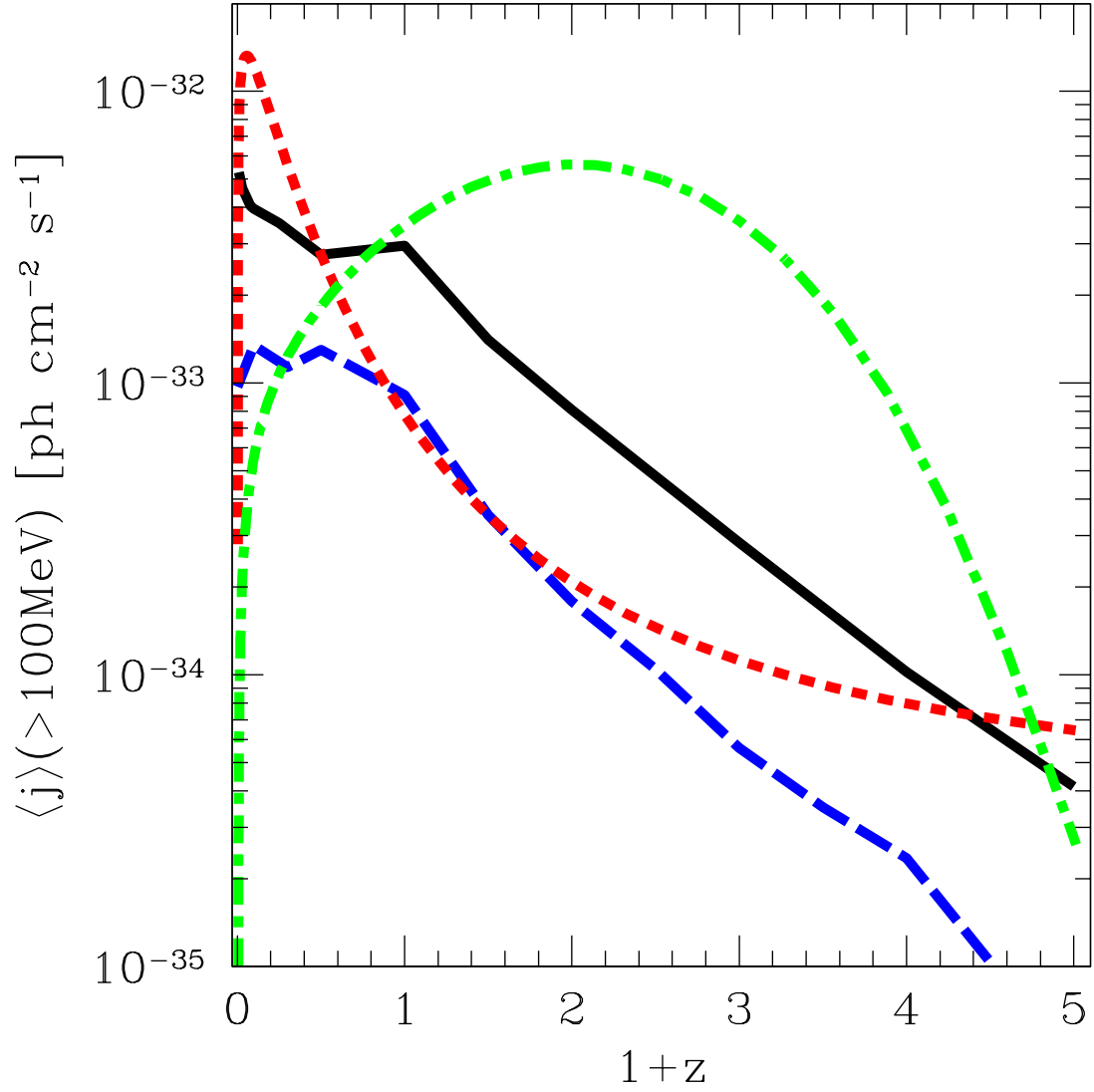}
\caption{Comoving volume-average emissivity as a function of redshift
  for IC emission from structure shocks (solid), $\pi^0$-decay and
  associated e$^\pm$ IC emission (long-dash), LDDE blazar model 
  (short-dash), PLE blazar model (long-short-dash).}
\label{j:fig}
\end{figure}

\begin{figure}
\plotone{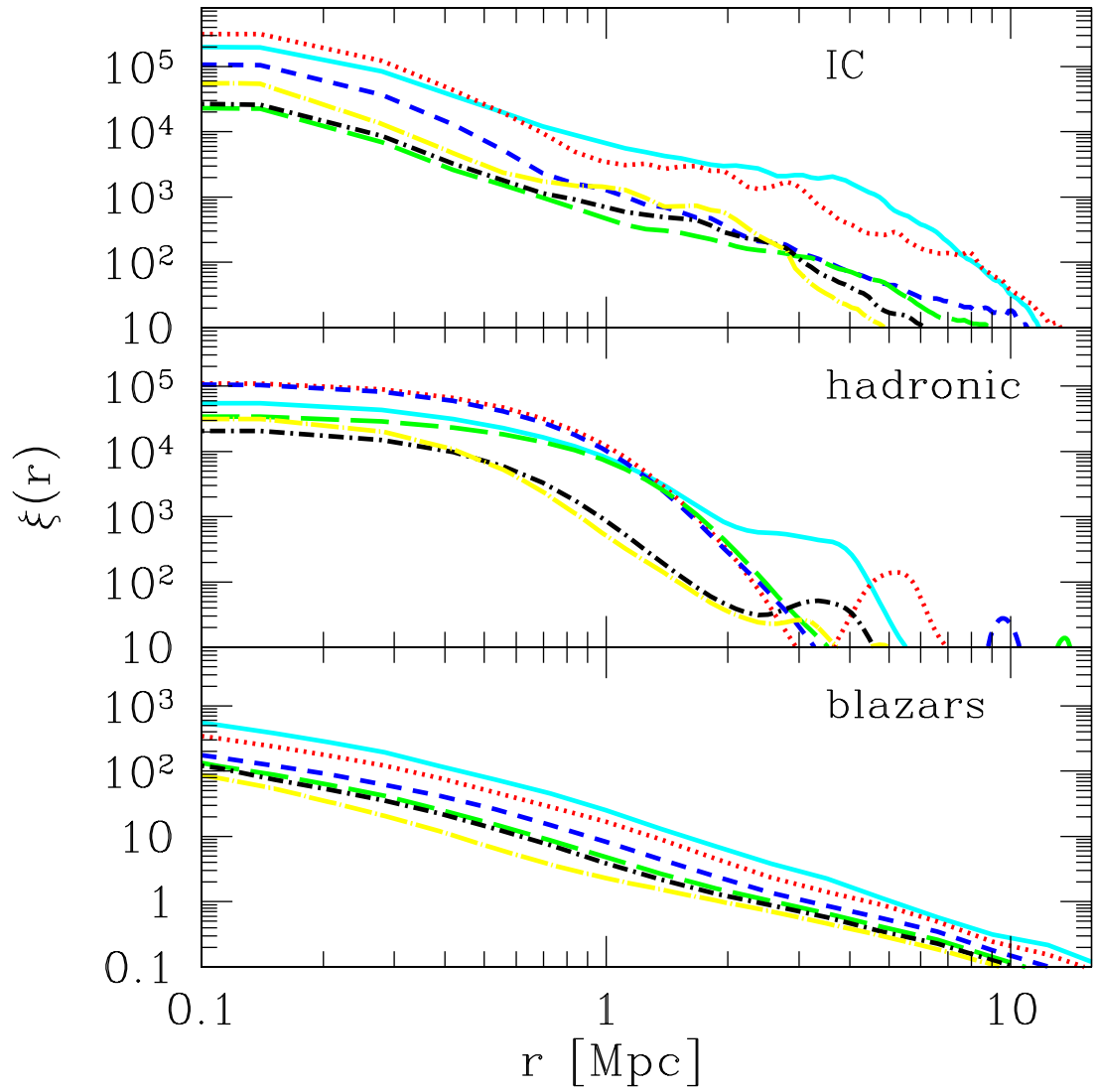}
\caption{Spatial correlation function.  Solid, dot, short-dash, 
  long-dash, dot-short-dash and dot-long-dash indicate, respectively,
  redshift $z$=$0.01,0.1,0.5,1,2,3$ for IC emission from structure
  shocks (top), $z$=$0,0.1,0.5,1,2,3$ for hadronic emission (middle)
  and $z$=$0.1,0.4,1,1.6,2,3$ (bottom) for blazars.}
\label{xi:fig}
\end{figure}

The redshift evolution of the average emissivity for different sources
is shown in Fig.~\ref{j:fig}.  With the chosen acceleration
efficiencies, IC (solid) and hadronic (long-dash) emission contribute
about 20$\%$ and 7$\%$ of the CGB, respectively, most of which will
likely remain unresolved.  While the hadronic emission originates
mostly in cluster cores, the IC emission is equally distributed in
shocks around clusters and filaments~\citep{min02}.  This explains the
different curve normalizations, despite the fact that the cluster
emission from both processes is below the EGRET upper limits.  The
unresolved blazars emissivity is also plotted for the LDDE (short-dash) 
and PLE (dot-long-dash) models, contributing a fraction about
30\% and 20\%, respectively, of the CGB. Note the marked difference
between the two models, with most of the contribution arising below
and above $z\sim 0.5$, respectively.
\placefigure{j:fig}
\placefigure{xi:fig}
The correlation function at various redshifts is shown in
Fig.~\ref{xi:fig}, for the IC (top), hadronic (middle) and blazars
(bottom) case.  Note the large difference in amplitude for the
correlation function of different potential CGB sources, especially at
distances of several Mpc. Note also both the power-law shape and
normalization change with redshift of the correlation function of
blazars, reflecting the underlying correlation of the host galaxies.

\begin{figure}
\plotone{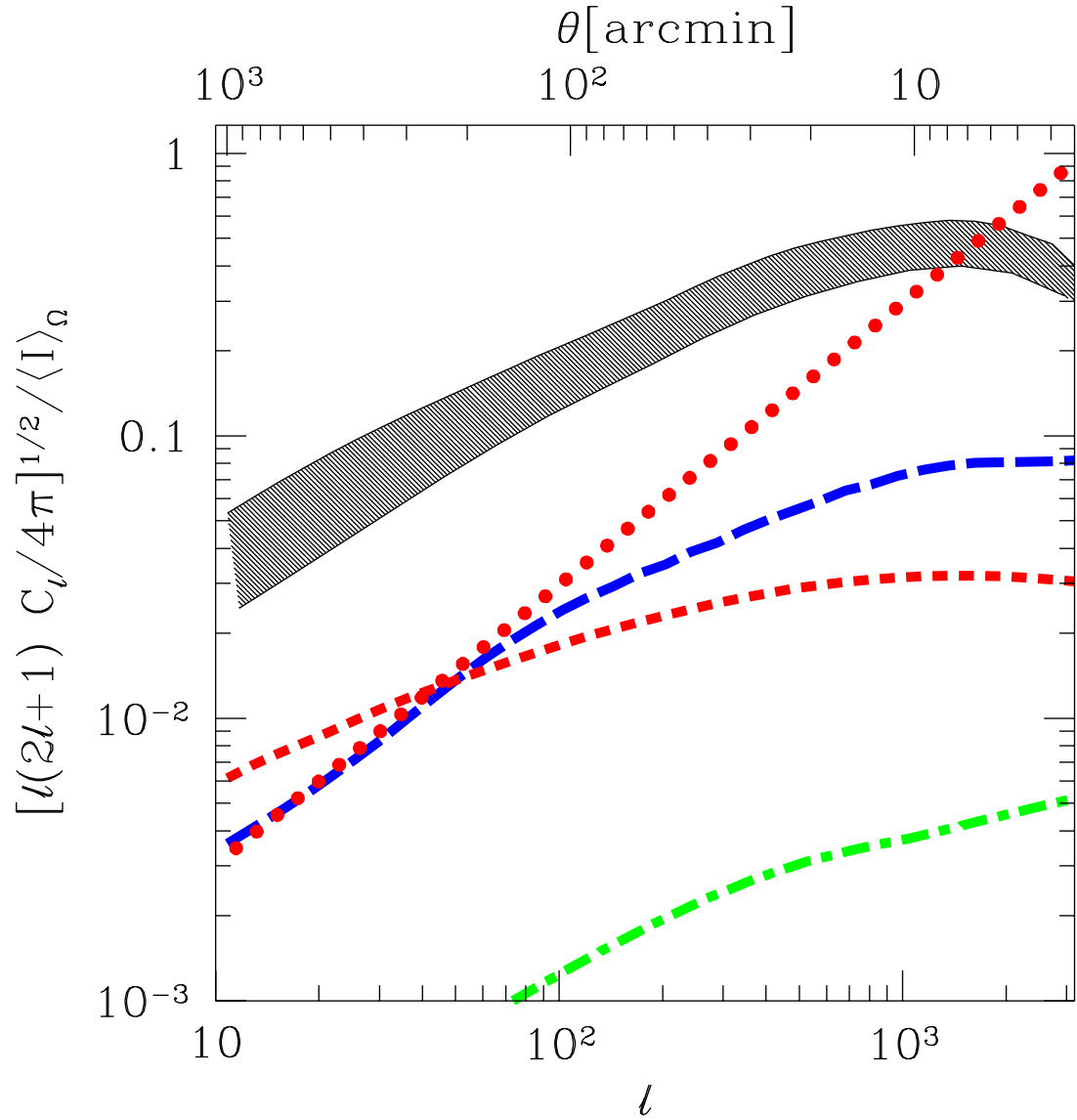}
\caption{Power spectrum of angular correlation function for: \gr
  emission from IC at structure shocks (shaded area), 
  hadronic processes in cluster cores
  (long-dash), LDDE blazar model (short-dash) and PLE blazar model
  (short-long-dash) and Poisson noise from unresolved blazars (dot).}
\label{cl:fig}
\end{figure}

Fig.~\ref{cl:fig} shows the power spectrum of fluctuations
of the integral photon intensity above 100 MeV. There
$\delta I_\ell/\langle I\rangle_\Omega 
=[\ell(\ell+1)C_\ell/4\pi]^{1/2}/\langle I\rangle_\Omega$, 
is the logarithmic contribution to the
relative intensity variation by a multipole $\ell$, corresponding to
an angular scale $\theta \simeq 180^\circ/\ell$.  The strongest signal
is due to IC from structure formation shocks (shaded area).
This is due to the much stronger correlation of shocks compared to
blazars and their higher emissivity compared to hadronic processes
(cf. Fig.~\ref{j:fig} and~\ref{xi:fig}).
At $\ell\le 100$, or $\theta\simeq 1^\circ.8$, the fluctuations
produced in the IC emission scenario are about 10\% of the average
intensity and significantly above the Poisson noise from potential
unresolved blazars, which is very similar in both the PLE
and LDDE models (dot curve).  Such fluctuations should be measurable
by GLAST, given its angular resolution of 3--4$^\circ$ at 100MeV. The
fluctuations increase at smaller angular separations, $\ell\ge 100$,
still accessible by GLAST at higher energies, e.g. a few GeV, where
the angular resolution is $\sim 0.^\circ 5$.  Note that our results
for the relative fluctuations are qualitatively consistent with, albeit 
a factor two lower than, those in~\cite{walo00}.
Also, the fluctuations produced by hadronic emission (long-dash) are
smaller than those due to IC emission in direct proportion to the
ratio of the emissivities of the two processes.

For blazars, in both LDDE and PLE scenarios the Poisson noise
dominates the signal for $\ell \ge 50$. Appreciable signal from
spatial clustering of the sources is predicted only in the LDDE
scenario (short-dash) at $\ell \leq 50$. At $\ell \sim 50$, accessible
by GLAST at 100 MeV, the intensity fluctuations are expected to be at
the level of a few per cent, well below the signal from structure
shocks.  In the PLE model the angular fluctuations due to spatial
clustering are much lower and always below the Poisson noise. The
reason is that in this model most of the unresolved emission is
produced at high redshifts, where the amplitude of the correlation
function decreases (cf Fig.~\ref{xi:fig}) and a correlated region
appears projected on smaller angular scales in the sky (at least up to
$z\leq$2 for the assumed cosmological model).  This, however, does not
affect the Poisson noise which only depends upon the number density
and luminosity of the sources.

\placefigure{cl:fig}

Note that the CGB integral intensity is $I(\varepsilon)\propto
\varepsilon^{-\alpha_I},~\alpha_I\ge 1$.  The spectrum of IC emission
from structure shocks is at least as flat as that~\citep{min02}, so
that relative to the average intensity, the intensity and fluctuations
contributed by this process should remain at least constant as a
function of photon energy.  In addition, the integral sensitivity of
GLAST up to a few GeV also scales as $\propto \varepsilon ^{-1}$ as is
the integral spectrum of blazars.  This implies that the number of
resolved sources should not change appreciably as a function energy
and that the Poisson noise from unresolved sources should also scale
with photon energy as the background intensity.  Therefore, the power
spectrum predicted in Fig.~\ref{cl:fig} should be roughly independent
of photon energy, between 100 MeV and a few GeV.  This implies that by
using information at different energies GLAST should be able to probe
the power spectrum of angular fluctuations on a significant range of
scales, from $\ell\simeq$ a few tens at 100 MeV and up to a
$\ell\simeq$ a few hundreds at 2 Gev.

Various sources of uncertainty affect the results presented in
Fig.~\ref{cl:fig}. Given an observed average CGB, $\langle I
\rangle_\Omega$, the intensity fluctuations $\delta I_\ell$ predicted
for the IC emission from structure shocks are directly proportional to
the assumed efficiency of CR acceleration. This parameter is highly
uncertain and an efficiency lower by an order of magnitude would
render the IC and LDDE model predictions indistinguishable. However,
in this case the IC model for the CGB would be ruled out, as it would
produce less than 2\% of it. The important point is that if structure
shocks contribute significantly to the CGB, GLAST direct observations
of nearby GCs~\citep{min02,min03} would determine the efficiency
parameter within a factor a few of the value assumed here.  The
predicted large intensity fluctuations would then provide a signature
of the (much larger) unresolved IC emission from structure shocks.
The consistency between observed \gr emission from GCs and CGB angular
fluctuations provides a test for the IC origin of the CGB.  At the
adopted numerical resolution the structure of strong shocks, where the
IC emission is produced, should have numerically
converged~\citep{ryuetal03}.  Nevertheless an uncertainty, $\Delta
{\cal M}$, in the determination of the shock Mach number, ${\cal M}$,
changes the log-slope $q$ of the CR distribution function by $\Delta
q=-\frac{1}{2}\frac{q^2}{{\cal M}^3}\Delta M$.  The shaded area in
Fig.~\ref{cl:fig} includes the range of fluctuations obtained for a
$\Delta {\cal M}$ due to a change in the postshock pressure $\Delta
P_2/P_2 \pm 30\%$. It also includes variations obtained when
changing $z_{min}$ from 0.05 to 0.01, which are negligible for
$\ell>$10$^2$ and lower the signal by $\sim$ 30\% at $\ell\sim10$.
Finally, we find that the occurrence of strong mergers adds a sampling
variance on the correlation functions, causing a factor $~1.5$
uncertainty (reduceable with a larger simulation box) in the predicted
fluctuations.  The ambiguity in the predicted blazars contribution to
the level of CGB fluctuations at large scales is represented by the
two curves for the LDDE and PLE models.  However, this will be largely
improved as GLAST which will discriminate between the two models based
on their very distinct predictions for the blazar luminosity function
and redshift evolution~\citep{nato06}.

We conclude that measuring the power spectrum of intensity
fluctuations together with the faint end of the blazar luminosity
function and \gr emission from GCs, should provide a valuable test of
consistency for the scenario in which the CGB is produced by either IC
emission at structure formation shocks or unresolved blazars.

~

We acknowledge useful comments from A. Pillepich.  FM
acknowledges support by the Swiss Institute of Technology through a
Zwicky Prize Fellowship.  Work at LANL was carried out under the
auspices of the NNSA of the U.S. DoE under Contract No.
DE-AC52-06NA25396.



\clearpage

\clearpage

\clearpage

\end{document}